\documentclass[11pt]{article}

\usepackage[T1]{fontenc}
\usepackage[utf8]{inputenc}
\usepackage{lmodern}

\usepackage[margin=0.7in]{geometry}

\usepackage{amsmath,amssymb,amsfonts}

\usepackage{graphicx}
\usepackage{xcolor}

\usepackage[numbers,sort&compress]{natbib}
\usepackage[colorlinks=true,
            linkcolor=blue,
            citecolor=blue,
            urlcolor=blue]{hyperref}

\usepackage{enumitem}
\usepackage{booktabs}
\usepackage{microtype}

\newcommand{\defemph}[1]{\textsc{#1}}
\newcommand{\fref}[1]{Fig.~\ref{#1}}

\title{Active Sensing Subserves Task-Level Control}

\author{
Andrew Lamperski$^{1}$,
Debojyoti Biswas$^{2}$,
Eric S.~Fortune$^{3}$,
John Guckenheimer$^{4}$,\\
Kathleen Hoffman$^{5}$,
and Noah J.~Cowan$^{2,6,*}$
\\[1em]
\small $^{1}$Department of Electrical and Computer Engineering, University of Minnesota \\
\small $^{2}$Laboratory for Computational Sensing and Robotics, Johns Hopkins University \\
\small $^{3}$Federated Department of Biological Sciences, New Jersey Institute of Technology \\
\small $^{4}$Department of Mathematics, Cornell University \\
\small $^{5}$Department of Mathematics and Statistics, University of Maryland, Baltimore County \\
\small $^{6}$Department of Mechanical Engineering, Johns Hopkins University\\
\small $^{*}$Correspondence: \href{mailto:ncowan@jhu.edu}{ncowan@jhu.edu}
}

\date{}

\begin{document}

\maketitle

\begin{abstract}
  Active sensing is traditionally defined as the expenditure of energy, typically in the form of movement, for obtaining information. Here, we propose that the combination of reliance on adaptive sensors, the linkage between movement and sensing, and task-level control inevitably gives rise to the emergence of active sensing movements. In this way, active sensing is not driven by sensory goals, such as minimizing uncertainty about the state, but rather is necessary for task-level control. This hypothesis, that active sensing subserves control, is supported by both empirical data from organisms and mathematical theory.  Interestingly, active sensing behaviors often occur in discrete epochs, interspersed with goal-oriented behavior. This suggests that animals switch between two behavioral modes with distinct control policies, an `explore' mode in which animals produce dynamic movements to shape sensory feedback, and an `exploit' mode in which animals produce slower compensatory movements that are directly related to achieving task goals. This strategy for feedback control that relies on adaptive sensors, active sensing, and mode switching is not commonly used in engineered systems despite being ubiquitous in biology. Engineered systems comprising state-of-the-art sensors, actuators, and mechanical designs can outperform animals with respect to ``cost functions'' such as maximum force generation, precision, and speed. Nevertheless, animals routinely achieve robust, graceful behaviors that are currently unmatched by engineered systems, suggesting that current control systems are insufficient. These insights, expressed in the language of control theory, may be critical for improving robotic sensing and control. 

\end{abstract}

\section{Introduction}

Active sensing is often described as ``movement for the purpose of sensing.'' This description implies that such movements are directed towards a set of \emph{sensory} goals, i.e.\ minimizing uncertainty about the organism's state or its environment, or minimizing sensory prediction error \citep{yang2016theoretical,friston2010free}. Recently, we developed a complementary perspective: both empirical evidence \citep{biswas2023mode-switching} and theory \citep{biswasexact2025} suggest that active sensing in biological systems emerges as a direct consequence of \emph{task-related feedback control}.

Organisms rely on sensors embedded in the individual for task control. Biological sensors exhibit a critical dynamic property known as \defemph{adaptive sensing}. Adaptive sensors exhibit robust responses to small \emph{changes} in sensory stimuli, while rejecting a constant, or nearly constant, baseline. Adaptive sensing is advantageous in that biological sensors achieve sensitivity to stimuli over wide dynamic ranges \citep{wandell1995foundations,barlow1961possible}. However, adaptive sensing can be disadvantageous in that it reduces \defemph{observability},  an important feature of a dynamical system that determines whether it is possible to infer the system's state from measurements. Adaptive sensing and other impediments, such as geometric constraints on sensing volumes \citep{snyder2007omnidirectional,nelson2006sensory,hartmann2001active,soatto2013actionable}, limit the observability of variables necessary to achieve task goals. This lack of observability can preclude stable closed-loop control of task-related behavior \citep{biswasexact2025}.

Here, we argue that the production of active sensing movements not only solves the problem of observability, but is an \emph{inescapable} consequence of feedback control of task-oriented behaviors in biological systems. We find that, for some systems, stability is impossible without active sensing movements when they rely on adaptive sensors. Specifically, the combination of (1) goal-directed behavior, (2) closed-loop control, and (3) geometric and dynamic sensory constraints lead to the emergence of self-generated movements that do not directly achieve a task goal, but nevertheless enable or improve task performance, i.e., \ active sensing emerges from task control. 

This finding suggests that the current understanding of active sensing movements, which emphasizes their role in achieving \emph{information goals} such as reducing uncertainty, gaining information \citep{yang2016theoretical}, or minimizing ``surprise'' \citep{friston2010free}---while certainly not incorrect---is an incomplete view.  In our framework, active sensing movements can also arise directly from the \textit{requirements of task-level feedback control.}  In other words, we argue that active sensing movements can emerge from a \defemph{control objective}, with no need for imposing information costs or goals.

\begin{figure}[t!]

\setlength{\fboxsep}{6pt}
\setlength{\fboxrule}{0.4pt}

\fbox{
\begin{minipage}{0.95\columnwidth}
\footnotesize

\textbf{Glossary}

\vspace{0.5em}

\textbf{\defemph{Adaptive sensing}}\\
A property of biological sensory systems in which sensors respond strongly to changes in stimuli while rejecting constant or slowly varying baselines, allowing sensitivity across wide dynamic ranges \citep{nelson1999prey,wark2007sensory}.

\vspace{0.4em}

\textbf{\defemph{Control objective}}\\
The task-level goal that drives behavior or movement generation in a control system. In this manuscript, active sensing movements are proposed to emerge naturally from the control objective itself \citep{todorov2004optimality,astrom1970stochastic}.

\vspace{0.4em}

\textbf{\defemph{Explore mode}}\\
A behavioral mode in which an organism generates active sensing movements to gather information about the environment or improve state estimation \citep{yang2016theoretical}.

\vspace{0.4em}

\textbf{\defemph{Exploit mode}}\\
A behavioral mode in which movements are generated primarily to achieve the task goal directly, such as minimizing tracking error \citep{todorov2002optimal}.

\vspace{0.4em}

\textbf{\defemph{Mode-switching}}\\
A strategy in which a system alternates rapidly between explore and exploit modes. Such switching strategies can outperform classical persistent excitation approaches in adaptive estimation and control \citep{mehra1974optimal,ljung1999system}.

\vspace{0.4em}

\textbf{\defemph{Non-convex}}\\
An optimization problem is convex if its cost function $f$ has the property $f(\lambda x_1+(1-\lambda)x_2)) \leq \lambda f(x_1) + (1-\lambda)f(x_2)$ for all $x_1$ and $x_2$ and any $\lambda\in[0,1]$. A local minimum of a convex optimization problem is always a global minimum. Non-convex optimization problems may contain multiple local optima, making them difficult to solve \citep{boyd2004convex}.

\vspace{0.4em}

\textbf{\defemph{Normative theories}}\\
Theoretical frameworks that compare observed behavior against mathematically optimal strategies, often using optimization-based principles such as optimal control \citep{todorov2004optimality} 
\vspace{0.4em}

\textbf{\defemph{Observability}}\\
A property of a dynamical system describing whether the internal state of the system can be inferred from available measurements \citep{kalman1960contributions,chen1999linear}.

\vspace{0.4em}

\textbf{\defemph{Persistent excitation}}\\
A classical control-theoretic strategy in which sufficiently rich or varying inputs are applied to ensure that system parameters or states remain identifiable \citep{ljung1999system,sastry2011adaptive}.

\vspace{0.4em}

\textbf{\defemph{Plant}}\\
In control theory, the physical system being controlled. In this manuscript, the locomotor dynamics of the organism are treated as the plant \citep{astrom1970stochastic}.

\vspace{0.4em}

\textbf{\defemph{Separation principle}}\\
A standard control-system design methodology in which state estimation and control are treated as separate problems: measurements are used to estimate system state, and the estimate is then used to compute control actions \citep{stengel1994optimal,kalman1960contributions}.

\vspace{0.4em}

\textbf{\defemph{Template model}}\\
A simplified mathematical representation of a system that captures the essential dynamics needed for analysis and control design. Template models are commonly used in biomechanics and legged locomotion research \citep{full1999templates}.

\end{minipage}
}

\end{figure}

Nevertheless, showing that active sensing can arise from closed-loop control alone does not by itself reveal the computational strategies the nervous system might use to generate those movements in service of task-level goals.
In fact, as we discuss later, when feedback control is posed as an optimization problem, the cost function is typically \defemph{non-convex}, and solutions require heuristic strategies \citep{biswas2018closed}. 

Thus, we turned to nature for insights and discovered that animals produce active sensing movements in discrete epochs. We termed these epochs of active sensing movements as \defemph{explore mode}. Animals also produce epochs in which they generate movements to directly achieve task goals, such as minimizing tracking error, which we termed \defemph{exploit mode}. We found that animals across a large taxonomic diversity rapidly switch between explore and exploit modes. Interestingly, this \defemph{mode-switching} strategy outperforms a classical method from control theory based on generating \defemph{persistent excitation}, achieving lower tracking error and reduced control effort in simulations \citep{biswas2023mode-switching}.

This approach---namely, the use of adaptive sensors, active sensing, and mode-switching---is almost the exact opposite of what is typically found in engineered feedback control systems. Engineers rely on what is commonly known as the \defemph{separation principle}, a design methodology in which the system uses sensory feedback to build a state estimate, and this estimate is used to guide control actions. Importantly, this principle is optimal in certain cases (such as the linear, time-invariant case), but as we showed \citep{kunapareddy2018recovering,biswasexact2025} and will review below, it does not work for the case of adaptive sensors---the most common type in nature. 

Machines are typically designed from the ``ground up'' with sensors that are mathematically well-characterized and provide ``absolute'' sensing (i.e.,\ non-adaptive). Despite the fact that engineered systems benefit from actuators and sensors that are more powerful and precise than those found in biology, animals nevertheless still dramatically outperform engineered systems in terms of task-level control and robustness. We assert that adaptive sensors, active sensing, and mode switching are likely to improve control of engineered systems and should be part of a robot designer's toolkit. This commentary is focused on showing how studying nature's strategies for control-driven active sensing can inspire a new paradigm for robotic active sensing. That said, the performance gap between animals and machines is not just a matter of control algorithms. Biomechanics and morphological intelligence---the body’s passive dynamics, compliant structures, and physical interactions with the environment---can play at least as important a role as neural control in producing robust behavior \citep{pfeifer2007self,full1999templates}. 

\section{Bio-inspiration: Weakly Electric Fish as a Model System for Active Sensing}

\begin{figure*}[tbh!]
  \centering
  \includegraphics[width=0.90\textwidth]{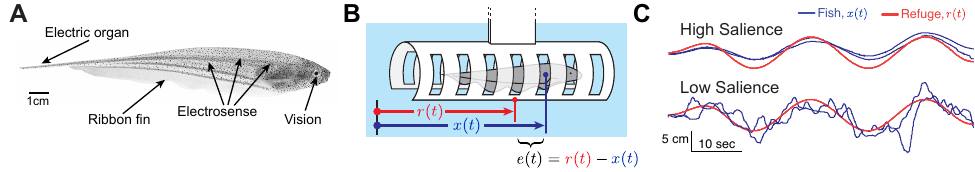}
  \caption{\textbf{The weakly electric glass knifefish, \emph{Eigenmannia
  virescens}, as a model for active sensing.}
  \textbf{(A)} This species has independent image-forming sensory systems,
  vision and electrosense, and a unique locomotor plant, a ventral ribbon fin
  that allows them to swim nearly equally well forwards and backwards.
    \textbf{(B)} The fish exhibit a natural behavior in which they maintain
    their position inside a longitudinally moving refuge. The refuge motion
    ($r(t)$) and fish movement ($x(t)$) do not exactly match, resulting in
    a dynamic sensory slip or ``error signal'', $e(t) = r(t)-x(t)$.
    \textbf{(C)} \emph{Eigenmannia} (two representative trials are shown in blue for each sensory-salience condition) track a moving refuge (red curves) under both high sensory salience (high luminance; vision and electroreception) and low sensory salience (low luminance; electroreception only). Under low-salience conditions, fish generate larger active swimming movements superimposed on the tracking behavior, consistent with enhanced active sensing. 
    }
\label{fig:fish}
\end{figure*}

The insights that active sensing is an inevitable consequence of task-oriented feedback control using adaptive sensors, and the widespread use of mode-switching as an active sensing strategy in biology, emerged from an unusual and fortuitous choice of animal model system: refuge tracking in the weakly electric glass knifefish \emph{Eigenmannia} (\fref{fig:fish}A). Despite its unusual nature, this system has features that help overcome challenges that limit the discovery and application of bio-inspired solutions.

Bio-inspiration suffers from two fundamental features of biological systems. First, organisms evolve idiosyncratic solutions to functional challenges. Consider the myriad forms of terrestrial locomotion, from hexapods to quadrupeds to bipeds: species have specialized adaptations to common problems that are not typically generalizable. Second, animal systems are notoriously nonlinear even within their normal functional ranges, making modeling and analysis difficult.


At first glance, \emph{Eigenmannia} might appear to be a poor choice of animal model as it is defined by idiosyncratic features, not least of which is its ability to generate weak electric fields that are used in localizing nearby objects in complete darkness. This electric field is generated by stacks of modified, non-contractile muscle cells along the sides and in the tail of the animal \citep{caputi2023living}. The electric fields spread in the water surrounding the fish up to about a meter distance. These fields are detected by a class of adaptive sensors, known as tuberous electroreceptors, embedded in the skin of the animal \citep{heiligenberg1989coding}. These receptors detect changes in the electric field that occur as the field passes through or around nearby objects that have resistances and capacitances that differ from the water. The electroreceptor array across the body allows the animal to resolve spatial patterns of nearby objects \citep{heiligenberg1975theoretical,assad1999electric,babineau2007spatial}. 

On the other hand, the behavior we study, refuge tracking, and the undulating ventral ribbon fin that it uses to swim permit the application of linear tools for analysis. In refuge tracking, fish swim back and forth to maintain their position within a moving refuge \citep{bastian1982vision,rose1993longitudinal,rojas2002multisensory}. Critically, the stimulus and response, movement of the refuge and fish respectively, occur primarily in a \emph{single linear dimension} (\fref{fig:fish}B,C). Further, although the ventral ribbon fin is itself a supremely complex locomotor organ, its performance can be captured by a single measurement that has a linear relation to the movement of the fish. That measurement is the ``nodal point'' of the fin, where rostro-caudal traveling waves that emerge from the caudal and rostral ends of the fin meet \citep{sefati2013mutually,uyanikvariability2020}.

The consequence is that each component of the system can be captured as a simple linear system, including the locomotor \defemph{plant}. This includes a primary sensory system used in the control of this behavior, the electrosensory system. Because the electroreceptors are embedded in the skin of the fish, there is a perfect correlation between the position of the fish and the locations of its electroreceptors.

These simplifications effectively `expose' the control policy for analysis: exposed insofar as no other complex analyses were necessary to address the dynamics of the control system. The application of linear system identification of the animal's tracking behavior \citep{cowan2007critical,uyanikvariability2020}, for example, has revealed several interesting nonlinearities in task-level control \citep{roth2011stimulus,yangmodeling2025,yang2024sensorimotor}. Remarkably, our reanalysis of time series data from a spectrum of organisms across size, life history, task features, sensory modalities, and phylogenetic background \citep{biswas2023mode-switching} found similar properties to those we observed in our analysis of active sensing during refuge tracking in \emph{Eigenmannia} \citep{stamper2012active,biswas2018closed}.

\section{Control Policies in Engineering and Biological Systems Obey Common Fundamental Constraints}

For a given environment (i.e., physical plant and sensors), controller performance is limited by fundamental laws that are well-understood from control and estimation theory \citep{crassidis2011optimal,lewis2012optimal}. For example, for a given set of sensors, no estimation strategy, whether implemented by a computer or neurons, can outperform the Bayes filter \citep{crassidis2011optimal}.

For the study of active sensing, relevant fundamental constraints arise in a simplified computational model of weakly electric fish, devised by \cite{kunapareddy2018recovering}. 
This model is surprisingly simple. It is built around the simple second-order \defemph{template model} for the task plant \citep{sefati2013mutually,uyanikvariability2020}:
\begin{equation}
  \label{eq:taskplant}
  \dot x =
  \begin{bmatrix}
    0 & 1 \\
    0 & -b
  \end{bmatrix}
  \dot x +
  \begin{bmatrix}
    0 \\ k
  \end{bmatrix}
  u,
\end{equation}
where $x$ is the state vector comprising position ($x_1$) and velocity ($x_2$), $\dot x$ denotes the time derivative of the state vector, $u$ is the offset of the nodal point from its equilibrium position, $b$ is the linear drag term that emerges due to the counter-propagating wave dynamics, and $k$ is the gain factor relating nodal-point movement to force. For details, see \cite{sefati2013mutually}. For simplicity, we set $k = 1$.

\begin{figure*}[tbh!]
\includegraphics[width = \textwidth]{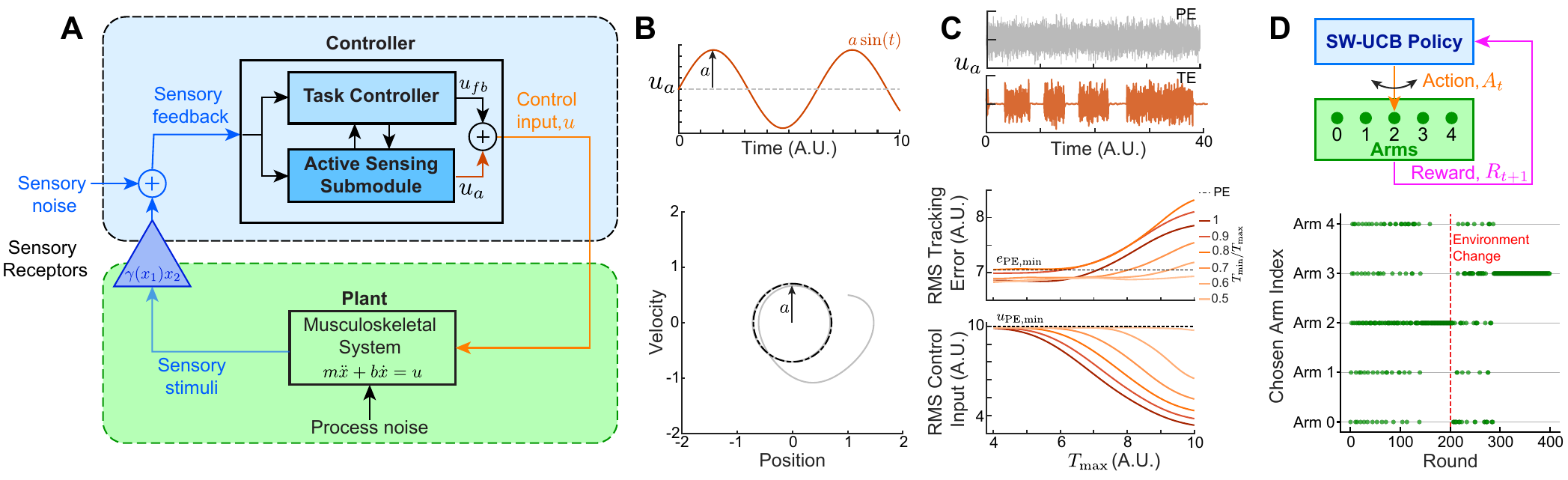}
\caption{\label{fig:theory} \textbf{Computational models of exploration in biological active sensing and reinforcement learning.}
\textbf{(A)} Schematic illustrating the proposed interaction between task-level control and active sensing, adapted from \citep{biswas2018closed,biswasexact2025}. Sensory feedback is processed jointly by a task controller ($u_{fb}$, standard state-feedback control law) and an active-sensing submodule (time varying, $u_a$), whose outputs combine to generate motor commands ($u=u_{fb}+u_a$) that drive the plant. Through movement, the active-sensing module modulates incoming sensory information and shapes the observability of the environment.
\textbf{(B)} Stabilization of the computational model in \citep{kunapareddy2018recovering,biswasexact2025} to a limit cycle using feedback control together with persistent sinusoidal active sensing. 
Top: sinusoidal active-sensing input. Bottom: phase portrait illustrating the resulting position--velocity trajectory over time.
\textbf{(C)} Comparison between a heuristic uncertainty-driven triggered switching strategy proposed in \citep{biswas2023mode-switching} and a conventional persistent-excitation strategy commonly used in adaptive control. Top: active-sensing signals for persistent (PE) and triggered excitation (TE) schemes. Bottom: comparison of tracking performance and control effort under varying uncertainty thresholds. For suitable choices of switching parameters (\(T_{\min}\) and \(T_{\max}\)), the switching controller achieves lower tracking error and reduced control effort compared with a standard persistence-of-excitation controller (\(e_{\mathrm{PE},\min}\) and \(u_{\mathrm{PE},\min}\)). Adapted from \citep{biswas2023mode-switching}.
\textbf{(D)} Simulation of a nonstationary multi-armed bandit problem solved using the sliding-window upper confidence bound (SW-UCB) algorithm of \citep{cheung2019learning}. The upper schematic illustrates adaptive selection among candidate sensing actions based on reward feedback, whereas the lower panel shows the arm selected at each round. Initially, the algorithm explores multiple arms before converging to Arm~2, which yields the highest average reward. At round~200, the reward distribution changes, increasing uncertainty and triggering renewed exploration. The controller subsequently adapts and converges to selecting Arm~3, which now gives the highest average reward. 
}
\end{figure*}

\cite{kunapareddy2018recovering} added to this model an adaptive sensor, i.e.\ a sensor that asymptotically blocks constant or persistent stimuli (\fref{fig:theory}A). The idea is that as the animal moves, a given sensory receptor experiences a position-dependent sensory stimulus, say $s(x_1)$, and we model the response of the receptor as the time derivative of this, namely
\begin{equation}
  \label{eq:sensormodel}
  y = \frac{d}{dt} s(x_1) = \frac{\partial s(x_1)}{\partial x_1}x_2 = \gamma(x_1)x_2,\quad \gamma:=\frac{\partial s(x_1)}{\partial x_1}.
\end{equation}

Regardless of the choice of $s(\cdot)$,
this system is locally, linearly \emph{unobservable} and it is \emph{impossible} for a control system to stabilize this system to an equilibrium point \citep{kunapareddy2018recovering,biswasexact2025}, but for essentially any reasonable choice of the function, $s$ (polynomial, sine wave, etc) the system can be stabilized around time varying movements ($u_a$), as depicted in \fref{fig:theory}B.

In this way, biological and engineering systems are fundamentally limited by \emph{observability of state variables} \citep{cellini2023empirical,cellini2024discovering, kalman1960contributions}. Even for the remarkably simple system in Eqs.~\eqref{eq:taskplant}-\eqref{eq:sensormodel}, observability is only maintained with active sensing motions. Critically, sensor information gathering was not included in a cost function or objective, but rather arose out of achieving a simple task goal: remain as close as possible to a desired position, $x_1^*$. To do this \emph{requires} active sensing. 

Another set of fundamental constraints arise from optimality principles. While we do not claim that biological systems optimize particular costs, optimal controllers define limits of achievable performance for any controller, whether engineered or evolved. In particular, when accurate state estimates are available, animal movements can closely resemble trajectories predicted from optimal control \citep{todorov2002optimal,wolpert2011principles}.

In optimal control, a common cost takes the form
\begin{equation}
\label{eq:cost}
\text{Cost} = \text{(error from goal)} + R \times \text{(control effort)},
\end{equation}
where $R$ is a weight on energetic expenditure \citep{todorov2002optimal,harris1998signal,diedrichsen2010coordination,carver2009optimal}. For a particular physical plant, optimal control strategies give the smallest possible value for the cost. The theoretical optimal values place fundamental constraints on achievable control performance, and these constraints hold for both biology and engineering. It should be noted that ``performance'' in this context is narrowly construed in reference to the engineering cost function at hand.

In most cases, there is no way to perfectly optimize the control and estimation strategies, but ``good'' solutions can emerge via adaptive processes, such as evolution \citep{parker1980optimality}, reinforcement learning \citep{sutton2018reinforcement,bertsekas2019reinforcement}, or repetitive movement practice \citep{todorov2002optimal}.  While these good strategies can never perform beyond the fundamental constraints of control, in trained movements, they can approach the limits \citep{harris1998signal,todorov2002optimal}. As a result, for many specific movement and estimation tasks, animal behaviors resemble theoretically optimal strategies. 

It should be noted that aside from simple special cases (such as the linear quadratic Gaussian problem), optimal controllers can only be computed, or even reasonably approximated, when the state is fully observable. So, movement systems examined in an optimal control framework typically exhibit high sensor coverage.  Much of the work on optimal control focuses on primate reaching, which utilizes numerous proprioceptors and vision to enable precise localization of the limbs \citep{todorov2002optimal,wolpert2011principles}.  

More broadly, beyond simple error minimization schemes described
below, most control methods require an estimate of the state. For a
fixed collection of sensors, fundamental limits on information
processing constrain the accuracy and precision of state estimates. A
common engineering strategy for improving estimates is to change the
sensing architecture. In contrast, organisms adeptly use active
sensing strategies to improve their estimates---a capability currently
lacking in engineering.

\section{Engineers Avoid the  Need for Active Sensing via Sensor Design}

 The dominant engineering control architectures either perform \emph{error minimization} or \emph{approximate state feedback}. In both cases, the sensing architecture is designed to avoid the need for active sensing. 

The simplest control systems aim to directly minimize the error between a sensor reading and its desired value. For example, a thermostat cools or heats a room to achieve a desired temperature. Proportional-integral-derivative (PID) control, which is the main control architecture for electric motors, specifies control inputs entirely in terms of the error between sensor readings and desired values \citep{franklin1994feedback}. In temperature or motor control, the variable being controlled is sensed directly, and the error between actual and desired values gives sufficient information for control.  

In more complex control systems, such as autonomous vehicles, direct minimization of sensor error is impossible, and control approaches are typically based on state feedback \citep{rajamani2011vehicle,lewis2012optimal}. To achieve accurate state estimates, engineers often design and deploy sensors as needed. For example, in an autonomous vehicle, cameras placed around the vehicle enable sensing in all directions without active sensing. 

In engineering, changing sensors and adding new sensors to achieve task goals are relatively common. In contrast, adding new sensors for a particular task in biological systems is subject to evolutionary constraints. For example, in engineered systems, adding new cameras at different locations on a device is a routine solution, while we find that vertebrates usually have two eyes (or three in animals with pineal eyes) positioned on the head, despite dramatic differences in behavioral repertoires across species.

In sharp contrast to an autonomous vehicle, human drivers with two eyes positioned in their heads rely heavily on actively moving their eye fixation locations in maneuvers such as lane changes.  The driver looks ahead to assess the distance to the vehicle in front, checks the mirrors to determine whether vehicles are approaching from behind, and then looks to the side to ensure that no vehicle is present in the blind spot.
These active sensing movements overcome sensor limitations. Below, we describe how animals use active sensing for control and then relate this to engineering principles.

\section{Bio-Inspired Active Sensing for Control}

\begin{figure*}[tbh!]
      \centering
      \includegraphics[width=0.8\textwidth]{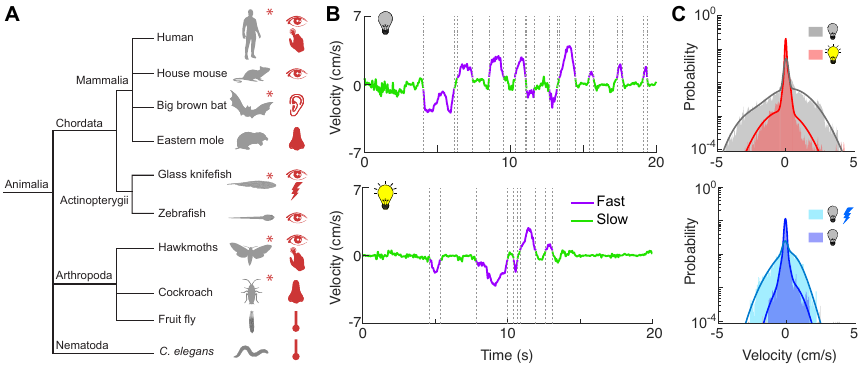}
      \caption{\textbf{Mode switching and sensory salience \citep{biswas2023mode-switching}.}
      \textbf{(A)} Mode switching during station-keeping tasks is found in species from humans to nematodes. The phylogeny highlights species in which mode switching has been quantified (grey) and the sensory modalities that mediate switching (red). Asterisks indicate species in which the effects of sensory salience are known. Adapted from \citep{biswas2023mode-switching}.
      \textbf{(B)} Two distinct behavioral modes, faster active sensing (purple) and slower task-control movements (green), during refuge tracking in weakly electric fish. In darkness (top), reduced sensory salience leads to significantly more transitions between modes (gray dashed lines) and increased time in the fast mode compared to light (bottom).
      \textbf{(C)} Top: decreased sensory salience (grey) increases high-velocity ``explore'' movements, evident as the wider high-velocity portion of the movement velocity histogram. Bottom: electrosensory jamming similarly reduces sensory salience and increases higher-velocity (light-blue) explore movements (data reanalyzed from \citep{chen2020tuning}). These patterns motivate computational, behavioral, and neural analyses of how transitions between modes are controlled and how sensorimotor control differs between them.}
      \label{fig:modeswitching}
\end{figure*}

We recently discovered that animals from amoeba to humans take advantage of the coupling between sensing and movement in two distinct behavioral modes. Across species, behavioral tasks, and sensing modalities, animals switch between `exploit' and `explore' modes (\fref{fig:modeswitching}A) \citep{biswas2023mode-switching}. In the exploit mode, animals produce slower compensatory movements that reduce their movement relative to features in the environment, thereby reducing the magnitude and rate of sensory feedback (\fref{fig:modeswitching}B). In the explore mode, animals produce faster movements, commonly known as `active sensing' movements, that increase the relative movement, thereby increasing the magnitude and rate of sensory feedback. These increases in relative movement have been shown to improve the animal's estimate of self-motion and environmental state. 

These modes appear as discrete epochs: animals alternate between bouts dominated by faster active sensing and bouts dominated by slower compensatory movements. Animals rapidly switch between modes, and the dynamics of switching depend on environmental conditions that affect sensory salience (\fref{fig:modeswitching}B,C). Critically, movement and sensing in both explore and exploit modes are under feedback control. 

Because switching itself is a stochastic control process, task performance may depend not only on how much time animals spend in each mode, but also on other characteristics of this process. In particular, the dwell times between switches could be constant (i.e., clock-driven), random, or triggered by some internal state. Similar to how dynamical systems theory \citep{MR1139515} uses normal forms to identify bifurcations in time series of deterministic systems,  we seek criteria to distinguish which simple stochastic models are consistent with observed statistics of mode switching in empirical time series.

The two modes have distinct control policies that nevertheless both rely on sensory feedback. To illustrate, consider stabilization tasks such as maintaining posture or position relative to environmental cues. During exploit mode, the goal of the animal is to minimize relative movement between the environment and itself. To achieve this goal, an animal generates compensatory movements to \emph{reduce} sensory signals generated by the relative movement between the animal and its environment. In contrast, during explore mode animals produce movements for sensing, i.e.,\ `active sensing.' Movements for active sensing typically \emph{increase} sensory signals via the generation of relative movement. In this case, the animal uses sensory feedback in opposing ways---to decrease the magnitude of sensory signals during exploit mode and to increase the magnitude during explore mode.

Mode switching may be mediated by internal representations of sensory uncertainty. These representations may, for example, include variance or entropy of a posterior distribution over task-relevant variables \citep{walkerstudying2023,koblingerrepresentations2021}. Our heuristic models suggest that when internal uncertainty exceeds a threshold, animals switch to an active sensing explore mode, which begins decreasing uncertainty by improving state (\fref{fig:theory}C). When uncertainty decreases to a sufficiently low value, animals return to goal-directed exploit mode \citep{biswas2023mode-switching}.
This perspective links mode switching to broader theories of explore--exploit tradeoffs in control and decision making.

The two modes are distinguished by substantial changes in control policy, although the specific behavioral manifestations vary across organisms and tasks. In many biological systems, these policy changes emerge through closed-loop interactions between sensing and movement, in which sensory feedback continuously shapes motor commands, and motor actions in turn reshape sensory feedback \citep{gibson1962observations,ahissar2003closed,ahissar2012seeing,zweifel2020defining,biswas2018closed}. Because sensing and action are tightly coupled, disentangling the respective contributions of sensory and motor systems is often challenging: closed-loop feedback can mask properties of the underlying subsystems.

Further, feedback is used in distinct ways during explore and exploit epochs. In refuge tracking, fish reduce the feedback error  $e(t)$ (Fig.~\ref{fig:fish}B) for exploit mode but increase $e(t)$ for explore mode. In other words, during exploit mode, sensory feedback is used to correct sensory slip---akin to a negative-feedback servo, an optimal controller \citep{todorov2002optimal}, or a proportional--integral--derivative controller \citep{cowantask-level2006,jmg5,roth2012task-level}. The goal is to minimize the error between a reference trajectory and the animal's state. However, during explore mode, motor commands are generated to create sensory slip \citep{gibson1962observations,stamper2012active,biswas2018closed,zweifel2020defining}. Movements are organized not primarily to reduce perceptual errors \citep{powers1973behavior}, but to enhance the structure, amplitude, or diversity of sensory inputs, thereby improving estimation of environmental or self-state \citep{bajcsy1988active,ahissar2016perception,biswas2018closed}. In summary, feedback is used in opposing ways between modes: in exploit mode, movements aim to eliminate sensory slip; in explore mode, they aim to generate sensory slip \citep{biswas2018closed}. The same physical circuit can therefore implement qualitatively different control policies depending on internal state and sensory context.

Mode switching between explore and exploit is ubiquitous across taxa and sensing modalities (\fref{fig:modeswitching}) \citep{biswas2023mode-switching}. Station-keeping behaviors such as quiescent stance \citep{kiemel2002multisensory,kiemel2011identification}, visual tracking \citep{hauperich2019makes,michaiel2020dynamics}, and electrosensory refuge tracking \citep{biswas2018closed,yang2024sensorimotor} all exhibit alternation between periods of high-velocity active sensing and low-velocity task control. These phenomena have implications for human balance and aging \citep{souza2024aging,cullen2023internal}.

How can we rationalize these explore--exploit movements for engineering design? Typically, 
\defemph{normative theories}
of active sensing seek strategies that trade off energetic cost, tracking error, and information gain \citep{yang2016theoretical,soatto2013actionable,chen2020tuning}. In contrast, active sensing movements can be required, even when information gain is not explicitly included as an objective.  
In the electric-fish-inspired theoretical model, fast active sensing movements improve observability and are required for stabilization. But what role does switching between explore (active-sensing) and exploit modes play?

It turns out that the phenomenon of exploration for task performance is common in reinforcement learning \citep{sutton2018reinforcement}. In particular, when uncertainty can grow over time, strategies that switch between exploration and exploitation can enhance task performance. Perhaps the most well-understood setting in which exploration/information subserves a goal-directed objective is the multi-armed bandit problem, which corresponds to a special type of reinforcement learning problem in which the state does not change and only rewards are observed (\fref{fig:theory}D). The classical setup models a row of slot machines that pay or take money according to unknown reward distributions. The rewards of a particular machine can only be observed by playing it (i.e., pulling its ``arm"). The goal is to accumulate the maximum amount of money by choosing a sequence of slot machines to play. To learn which machines pay the most, on average, they must be explored.

In this setting, the rewards are random, but their distributions are stationary over time. For this case, a variety of algorithms achieve near-optimal performance \citep{lattimore2020bandit}.  While the design details differ, they all begin with a phase of high exploration and transition to high exploitation as the uncertainty about the highest-paying machines decreases.

When the reward distributions can change over time, a more flexible strategy is needed to achieve near-optimal performance, namely switching between exploration and exploitation. The general scheme of these strategies is to run a variation of an algorithm designed for stationary rewards and then trigger a reset (and thus an exploratory phase) if uncertainty about the best decisions increases (\fref{fig:theory}B). This general approach to uncertainty-driven triggering exploration underlies a variety of provably near-optimal methods for bandit problems and more general reinforcement learning problems \citep{cheung2019learning,wei2021non,wang2025adaptivity}. The optimal attention switching strategy for binary decisions, which matches observations from human experiments, gives analogous rules for information gathering based on whether sufficient information for a decision has been found \citep{jang2021optimal}.
Note that for all of these optimization problems, there are no terms in the reward associated with information gathering, but all near-optimal methods utilize exploratory actions to gather the information required for the task.

The exploratory decisions of the multi-armed bandit can be viewed as analogous to the active sensing movements we observe in the electric fish. The work of \citep{biswas2023mode-switching} gives a heuristic uncertainty-triggered refuge-tracking strategy for a variation of the model from \citep{kunapareddy2018recovering} with noisy forcing, which simulates the effects of fluid forces on the fish. An internal estimator tracks the uncertainty of position; when uncertainty grows beyond a threshold, the controller enters an exploratory regime, generating high-velocity movements to increase electrosensory slip \citep{biswas2018closed} and reduce uncertainty. When uncertainty is low, the controller reverts to low-velocity exploit mode to maintain accurate tracking (\fref{fig:theory}C). In simulations, such strategies can outperform classical ``persistent excitation'' methods from adaptive control, with respect to a cost of the form in Eqn.~\eqref{eq:cost}  \citep{biswas2023mode-switching}. 

\section{Discussion}

\begin{figure}[tbh!]
      \centering
      \includegraphics[width=\columnwidth]{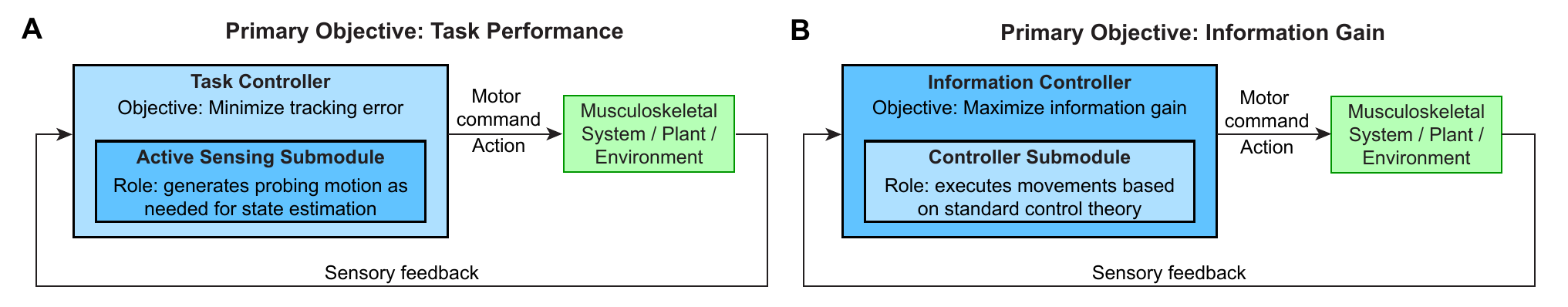}
      \caption{
\textbf{Comparison between active-sensing-for-control and infotaxis-style information maximization.}
In the active sensing framework, task control is primary: the controller minimizes tracking error, while an active sensing generator adds exploratory movements as needed to improve state estimation. In the infotaxis framework, information acquisition is primary: actions are selected to maximize expected information gain, often through a Bayesian belief state, and task execution is subordinate to information-seeking. 
Even within this framework, an additional lower-level active sensing \emph{sub-submodule} (not shown) may still be required to support state estimation for feedback control.}
\label{fig:active_sensing_vs_infotaxis}
\end{figure}

We are proposing a new framework for the emergence of movements for active sensing. In this framework, active sensing is an inevitable consequence of feedback control of behavior that relies on adaptive sensors embedded in an organism. This framework differs from previous thinking about movements for active sensing, which focus on how these movements are used to achieve sensory goals, such as minimizing uncertainty about the organism's state.  In contrast, our framework emphasizes that active sensing movements can emerge directly from the dynamical and observability constraints imposed by closed-loop feedback control with adaptive sensors. Critically, these two frameworks are not mutually exclusive: an animal could, for example, implement an `infotaxis' strategy for gathering information while also generating active sensing movements that emerge from its feedback control systems (\fref{fig:active_sensing_vs_infotaxis}).

The compatibility of these strategies can be illustrated via the meanings of the terms `explore' and `exploit.' Traditionally, these terms refer to how animals allocate time and effort in relation to tasks, such as foraging \citep{king2022optimal, werner1974optimal}. In this formulation, animals weigh benefits and risks in making decisions about how long to linger at a particular resource (exploit), versus when to move to find another (explore). These decisions are based on information available to the animal, including the value of the resource at hand and the potential distribution and value of other resources in the environment. Such strategies can be implemented in simple foraging systems found in single-celled organisms, or as cognitive processes found in more complex organisms, including humans \citep{addicott2017primer}. Our usage of the terms explore and exploit also involves the availability of information and the allocation of time and effort, but for the purpose of achieving task goals.

An animal may engage in both forms of explore and exploit as part of the same behavior. Consider an animal that is, from a foraging perspective, in exploit mode while eating a food resource. To achieve that task, an animal might employ behaviors that are under feedback control and rely on switching between explore and exploit movements, such as handling a prey item. In another example, obtaining information from the environment might be the task goal, which can lead to complex strategies for active sensing movements that depend on sensor volumes and the distribution of information in the environment \citep{chen2020tuning,vergassola2007infotaxis}. Again, the implementation of these strategies might be under feedback control that induces bouts of active sensing movements.

Beyond exploration, active sensing can also serve to improve the accuracy of spatial information. On a human scale, we experience the world around us as (almost) obeying Newton's laws of motion. Macroscopic objects---including ourselves---follow smooth paths. Our brains segment visual input into distinct objects and follow their motion. Many tasks (e.g., reading, catching a ball, or threading a needle) require high accuracy in locating and tracking objects. Cameras are a poor model for how our visual system achieves the needed accuracy. Our visual acuity depends upon eye movements that switch between \emph{saccades}, rapid ballistic rotations of the eyes, and fixation intervals in which the object of current interest is seen by the foveae, where the density of retinal photoreceptors is high \citep{rucci2015control}.

We believe that the implementation of active sensing and mode switching control policies in engineering design can significantly improve performance, especially within the realm of robotics. 
For example, trajectory planning for robots begins with computing a kinematically feasible trajectory. This strategy requires sufficient actuators to constrain the robot to this trajectory, sufficient sensors to observe the high-dimensional state space of the robot, and sufficient energy to execute the motion. Robots built this way are typically heavy, slow, and prone to falling. In contrast, humans walk in a dexterous manner while setting a seemingly small number of control variables consciously. Humans adaptively reset gains by initiating exploratory movements and sensing their effects. These effects are compared with a learned and continually updated \emph{efference copy} of predicted motion. Interestingly, robotic systems that embody this process via extensive training with volumes of motion capture data have produced dramatic improvements in performance.

\begin{figure*}
\centering
\includegraphics[width=\textwidth]{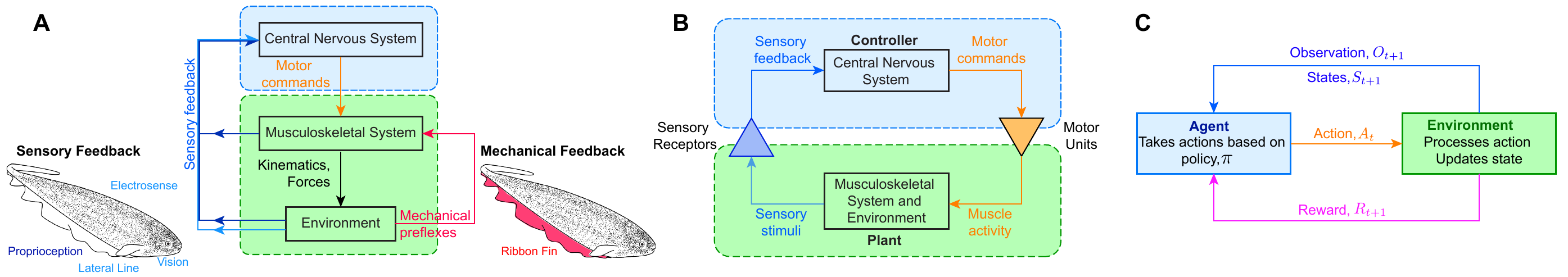}
\caption{
\textbf{Different perspectives on the same animal--environment interaction.} Colors denote corresponding elements across representations.
\textbf{(A)} Biological perspective. An animal (illustrated with a weakly electric fish) interacts with its environment through closed-loop sensorimotor dynamics. The central nervous system (CNS) generates motor commands that drive the musculoskeletal system, which in turn acts on the environment to produce motion, forces, and sensory stimuli. Multiple sensory modalities (e.g., electrosense, vision, proprioception, lateral line) provide feedback to the CNS, which updates motor output. In parallel, passive mechanical properties (preflexes) generate rapid responses without neural delay. Adapted from \citep{dickinson2000animals}.
\textbf{(B)} Control-theoretic perspective. The interaction is recast as a feedback control system in which the CNS acts as the controller and the combined musculoskeletal system and environment constitute the plant. Sensory pathways provide sensory feedback to the controller, while descending motor commands drive the plant through muscle activation. Because sensing occurs on the body, locomotion-induced interactions with the environment and passive mechanical feedback are embedded within the plant dynamics, reflecting the inseparability of sensing and actuation in embodied systems. Adapted from \citep{madhavsynergy2020}.
\textbf{(C)} Reinforcement learning perspective.
The interaction is interpreted as an agent--environment loop. The agent selects actions according to a policy, $\pi$, while the environment, representing the combined body and external world dynamics, evolves its state and returns observations and reward signals that guide subsequent actions. Here, reward provides an explicit objective signal that is implicit in the biological and control-theoretic formulations.
}
\label{fig:perspectives}
\end{figure*}

We end our commentary by presenting the emergence of mode switching from the perspectives of biology, control theory, and reinforcement learning, as illustrated in \fref{fig:perspectives}. In biological systems (\fref{fig:perspectives}A), behavior emerges from the interaction of the nervous system, musculoskeletal system, and environment \citep{dickinson2000animals}, and the details of these embodied dynamic interactions are critical for understanding the rules governing active sensing and mode-switching behavior. Control theory (\fref{fig:perspectives}B), when applied with careful attention to these biological details, can provide an abstracted framework that ``lumps'' the body--environment interactions into a plant model. In this view, there is a one-way flow of information between plant and controller, mediated by motor system outputs and sensory receptor inputs that serve as amplifiers \citep{del2009engineering}, facilitating the application of control-theoretic ideas \citep{madhavsynergy2020}. Finally, in reinforcement learning frameworks, the ``lumps'' differ from biological and control-theoretic perspectives: here the environment includes the dynamics of the organism and external world, returning state and reward signals that guide the agent's future actions (\fref{fig:perspectives}C), enabling us to draw on systematic analyses of the interplay between exploration and exploitation. In each framework, sensor limitations such as field of view constraints and adaptation to constant stimuli reduce observability, inevitably leading to the generation of active sensing movements to subserve control. The differences between the frameworks highlight the challenges for decoding the biological mechanisms for the generation of active sensing and mode switching, and the differences in approach for implementing these strategies in artificial systems.

\section*{Acknowledgments}

This work was supported by the Collaborative Research in Computational Neuroscience (CRCNS) program through NIH award R01NS147767 to N.J.C., E.S.F., A.L., and K.H.

\providecommand{\newblock}{}

\bibliography{jeb} 
\end{document}